\documentclass{pasj01}
\Received{$\langle$reception date$\rangle$}
\Accepted{$\langle$acception date$\rangle$}
\Published{$\langle$publication date$\rangle$}
\usepackage{url}
\usepackage{lineno}
\newcommand{\PKHY}[1]{{{#1}}}

\begin{document}

\title{Multiwavelength studies of G298.6$-$0.0: An old GeV supernova remnant interacting with molecular clouds}
\author{
Paul K. H. Yeung\altaffilmark{1,*},
Aya Bamba\altaffilmark{1,2},
Hidetoshi Sano\altaffilmark{3}
}
\altaffiltext{1}{Department of Physics, The University of Tokyo, 7-3-1 Hongo, Bunkyo-ku, Tokyo 113-0033, Japan}
\altaffiltext{2}{Research Center for the Early Universe, School of Science, The University of Tokyo, 7-3-1 Hongo, Bunkyo-ku, Tokyo 113-0033, Japan}
\altaffiltext{3}{Faculty of Engineering, Gifu University, 1-1 Yanagido, Gifu 501-1193, Japan}

\email{pkh.yeung@phys.s.u-tokyo.ac.jp}


\maketitle

\begin{abstract}

Hadronic $\gamma$-ray sources associated with supernova remnants (SNRs) can serve as stopwatches for the escape of cosmic rays from SNRs, which gradually develops from highest-energy particles to lowest-energy particles with time. In this work, we analyze the 13.7~yr \emph{Fermi}-LAT data to investigate the $\gamma$-ray feature in/around the SNR G298.6$-$0.0 region. With $\gamma$-ray spatial analyses, we detect three point-like components. Among them, Src-NE is at the eastern SNR shell, and Src-NW is adjacent to the western edge of this SNR. Src-NE and Src-NW demonstrate spectral breaks at energies around/below 1.8~GeV, suggesting an old SNR age of $>$10~kyr. We also look into the X-ray emission from the G298.6$-$0.0 region, with the Chandra-ACIS data. We detected an extended keV source having a centrally filled structure inside the radio shell. The X-ray spectra are well fit by a model which assumes a collisional ionisation equilibrium of the thermal plasma, further supporting an old SNR age. Based on our analyses of the NANTEN CO- and ATCA-Parkes HI-line data, we determined a kinematic distance of $\sim$10.1~kpc from us to G298.6$-$0.0.  This distance entails a large physical radius of the SNR of $\sim$15.5~pc, which is an additional evidence for an old age of $>$10~kyr. Besides, the CO data cube enables us to three-dimensionally locate the molecular clouds (MCs) which are potentially interacting with SNR G298.6$-$0.0 and could account for the hadronic $\gamma$-rays detected at Src-NE or Src-NW. Furthermore, the multiwavelength observational properties unanimously imply that the SNR--MC interaction occurs mainly in the northeast direction.

\end{abstract}

%%%%%%%\linenumbers

\section{Introduction}

Synchrotron X-rays \citep{Koyama1995} and GeV--TeV $\gamma$-rays (e.g., \cite{Aharonian2004, Acero2016}) from the regions of supernova remnants (SNRs) indicate that the shock fronts of SNRs are powerful sites of cosmic-ray acceleration. When a hadronic cosmic-ray particle (usually a proton or an atomic nucleus) from an SNR collides with gas in the interstellar medium, a neutral pion is produced and quickly decays into two $\gamma$-ray photons that we can observe. However, the escape of accelerated particles from SNR shocks remains an undetailed process.

$\gamma$-ray spatial analyses enable us to compare the $\gamma$-ray morphology of an SNR-associated source with the SNR morphologies in radio continuum and X-ray, as well as the molecular-cloud (MC) distributions traced by radio line emissions and the shock-cloud interaction sites  indicated by ``maser" spots. It is interesting to note that the $\gamma$-ray location of such a source can generally be offset from the SNR shell yet more consistent with a MC clump. The $\gamma$-ray emissions associated with RX J1713.7$-$3946 (Figure~1 of \cite{Fukui2012}), Kes 79 (Figure~2 of \cite{He2022}) and Kes 41 (Figure~1 of \cite{Liu2015}) are conspicuous examples of this phenomenon, which representatively agrees with the fact that pion-decays occur at proton collision sites (MCs) rather than the proton acceleration site (SNR shock). More accurately speaking, in a hadronic model for an SNR--MC system, the actual $\gamma$-ray emitters are the MCs impacted by the SNR-accelerated cosmic-ray protons, in addition to a radiative shell behind the SNR shock. We will also embrace this concept when interpreting the $\gamma$-rays from our targeted regions.

Spectral breaks in the GeV $\gamma$-ray band \citep{Acero2016} are common signatures of SNR--MC interaction systems observed by \emph{Fermi} Large Area Telescope (\emph{Fermi}-LAT), reflecting the escape of over-energetic particles from the vicinities of SNRs. More interestingly, there have been a number of attempts to formulate the relation of the $\gamma$-ray spectrum with the SNR age (e.g., \cite{Dermer2013, Bamba2016, Zeng2019, Suzuki2020, Suzuki2022}). All these studies point to a trend that, as an SNR grows older, the $\gamma$-ray spectrum of the SNR--MC source becomes softer and the spectral peak shifts to a lower energy. Such an observed trend is consistent with a theoretical prediction that the escape of cosmic rays from SNRs gradually develops from highest-energy particles to lowest-energy particles with time \citep{Ptuskin2003, Ptuskin2005}. Hence, SNR--MC interaction sources can serve as stopwatches for this escape process. In particular, it is important for us to investigate $\gamma$-rays from regions of old ($\gtrsim$10~kyr) SNRs more deeply, for the sake of figuring out the later evolution stages of the escape. 

SNR G298.6$-$0.0 was detected by Molonglo at 408~MHz and by MOST at 843~MHz, and was found to have a flat radio spectral index of -0.3 \citep{Shaver1970, Kesteven1987, Whiteoak1996}. \citet{Reach2006} claimed a possible detection of infrared emission from the direction of G298.6$-$0.0, suggesting an encounter between the SNR shock and a high-density medium. These conditions set up an adequate environment for emitting GeV $\gamma$-rays, which are also detected by \emph{Fermi}-LAT \citep{Acero2016}. \citet{Bamba2016} presented the first X-ray imaging spectroscopy for G298.6$-$0.0 using Suzaku data, and they characterise the X-rays from this SNR as thermal emissions of plasma in collisional ionisation equilibrium. In addition, \citet{Bamba2016} found a centrally filled X-ray structure inside the radio shell, categorising G298.6$-$0.0 as a mixed-morphology (i.e. thermal composite) SNR. These X-ray properties suggest that this SNR is relatively older and thus ideal for our studies. Intriguingly, there has not yet been a published estimation of the age of G298.6$-$0.0, encouraging us to provide the very first constraints on its age. 

To the southwest of G298.6$-$0.0 (about 0.3$^\circ$ apart), there is another SNR -- G298.5$-$0.3, detected by Molonglo and MOST at 408~MHz and 843~MHz respectively \citep{Shaver1970, Whiteoak1996}. A claimed possible detection of infrared emission from the direction of G298.5$-$0.3 \citep{Reach2006} makes it likely to be associated with a GeV $\gamma$-ray source as well. In view of the relatively poor angular resolution of \emph{Fermi}-LAT, the $\gamma$-rays at/around G298.5$-$0.3 contribute  non-negligible contamination to the measurements of $\gamma$-ray properties of our targeted G298.6$-$0.0.

This paper reports our $\gamma$-ray observational results on the regions of SNR G298.6$-$0.0, SNR G298.5$-$0.3 and their vicinities (in \S\ref{gammarayresults}). Our X-ray (\S\ref{Xrayresults}) and radio (emission lines; \S\ref{radioresults}) observational results on G298.6$-$0.0 are also presented in this paper. We interpret the multiwavelength properties of G298.6$-$0.0 comprehensively, and propose the hadronic scenarios of SNR--MC interaction (in \S\ref{hadronic_SNR--MC}). In turn, we estimate its age for the first time (in \S\ref{oldage}). The interpretations for G298.5$-$0.3 are beyond the scope of this paper.

\section{Observations \& data reductions}

\subsection{\emph{Fermi}-LAT $\gamma$-ray data}

In this work, we use the  Fermitools version 2.0.8  to reduce and analyze the \emph{Fermi}-LAT data. We select  Pass 8 (P8R3) `SOURCE' class events collected between August 4, 2008, and April 28, 2022. The region of interest (ROI) we choose  is $21^{\circ}\times21^{\circ}$ centered at RA =$12^{h}13^{m}40.80^{s}$, Dec=$-62^{\circ}37^{'}12.0^{''}$ (J2000). This ROI center is inside the radio dimension of SNR G298.6$-$0.0. 

We perform a series of binned maximum-likelihood analyses, with an angular bin size of 0.05$^\circ$ that is sufficiently small to well sample the point-spread function (PSF) at energies up to $\sim$20~GeV (see
SLAC\footnote{\label{slac}\emph{Fermi} LAT Performance:
\url{http://www.slac.stanford.edu/exp/glast/groups/canda/lat_Performance.htm}}). To better model the background, the Galactic diffuse component (gll\_iem\_v07.fits), the isotropic diffuse component (iso\_P8R3\_SOURCE\_V2\_v1.txt), and the sources in the \emph{Fermi} LAT 12-Year Point Source Catalog (4FGL-DR3; \cite{Abdollahi2022}) are included as background sources in our analyses.  We set free the spectral parameters of the sources within 5.5$^\circ$ from the ROI center (including the normalizations of the Galactic diffuse background and of the isotropic diffuse background) in each analysis. For the sources at angular separation beyond 5.5$^\circ$ from the ROI center, their spectral parameters are fixed to the catalog values. 

In the newest 4FGL-DR3 catalog, 4FGL J1213.3$-$6240e is  assigned with  a disk morphology enclosing SNR G298.6$-$0.0 and SNR G298.5$-$0.3. In this work, we decompose 4FGL J1213.3$-$6240e into a number of spatial components that we identify.

\subsection{Chandra-ACIS X-ray data}

With the aid of the CIAO version 4.14, we reprocess, reduce and analyse the Chandra-ACIS data of the observation ID 14889. This dataset contains the events in/around the dimension of SNR G298.6$-$0.0. The observation started at 2013 September 03 12:34:17. 

The data reprocessing is done with the CIAO command ``chandra\_repro" and the resulted exposure time is 20~ks. The command ``fluximage" is operated on the event file to generate count maps, exposure maps and flux maps (i.e. exposure-corrected maps). ``specextract" is implemented to extract spectra for source regions and background regions, respectively. Then, the Sherpa version 4.14.0 enables us to produce background-subtracted spectra and to perform spectral fittings on them.

\subsection{NANTEN CO- and ATCA-Parkes HI-line data}

In order to trace the spatial distribution of MCs and dense media at/around SNR G298.6$-$0.0, we reduce and analyse the $^{12}$CO($J$=1--0) (115~GHz) line data collected by NANTEN \citep{Mizuno2004} as well as the combined ATCA-Parkes data of HI (1.4~GHz) line emissions provided by the Southern Galactic Plane Survey (SGPS; \cite{McClureGriffiths2005}). The angular resolution is 156$^{''}$ for the NANTEN $^{12}$CO($J$=1--0) data and 130$^{''}$ for the ATCA-Parkes HI data. The typical noise fluctuation per velocity resolution of 1~km~s$^{-1}$ is $\sim$0.2~K for the NANTEN CO data and $\sim$1.4~K for the ATCA-Parkes HI data. Each data cube has three dimensions: the right ascension, the declination, and the radial velocity with respect to the local standard of rest ($V_\mathrm{LSR}$). 

\section{Results of data analyses}

\subsection{\emph{Fermi}-LAT results}\label{gammarayresults}

\subsubsection{$\gamma$-ray spatial morphology}

\begin{figure*}
  \begin{center}
  \includegraphics[width=55mm]{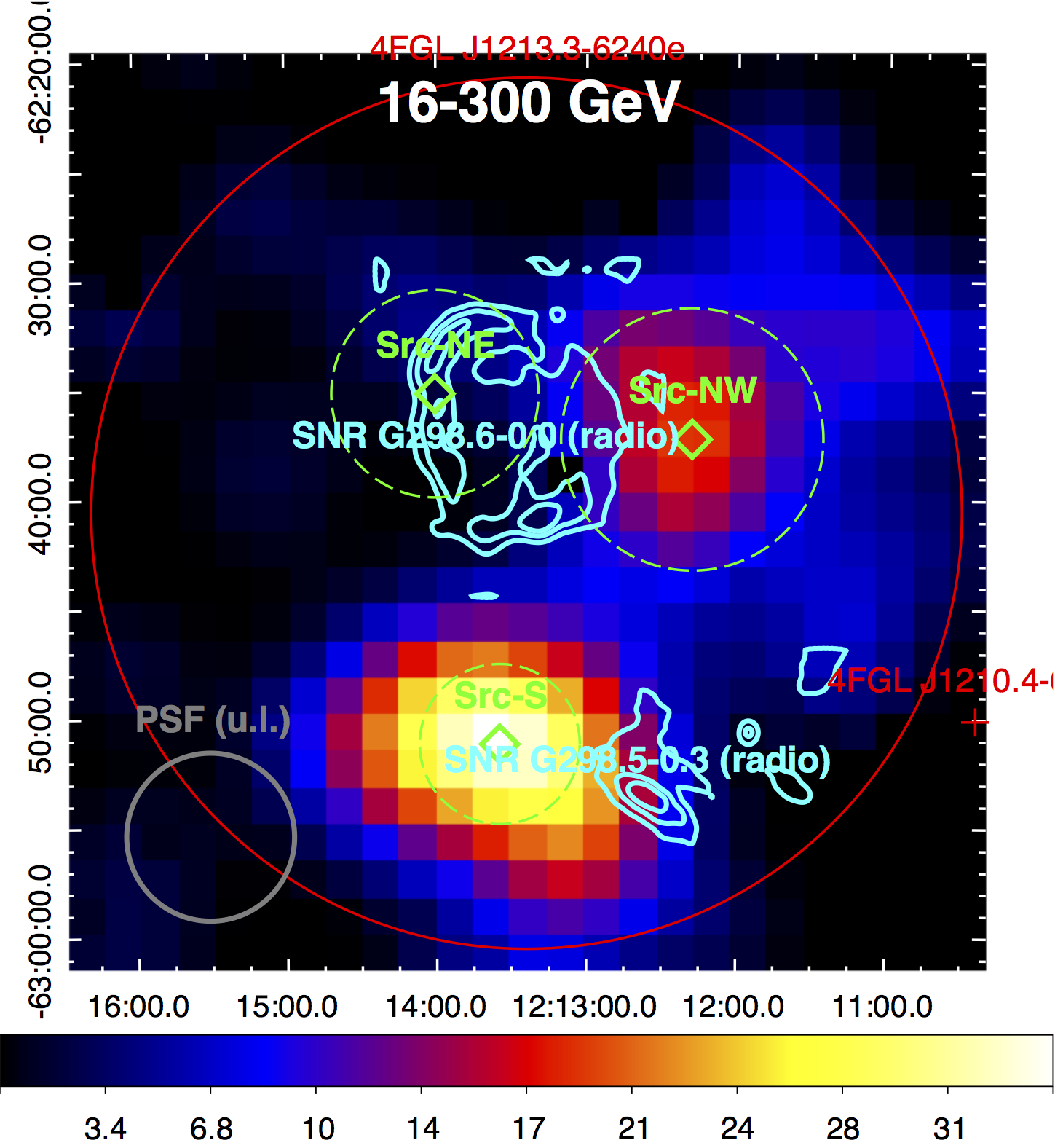}
  \includegraphics[width=55mm]{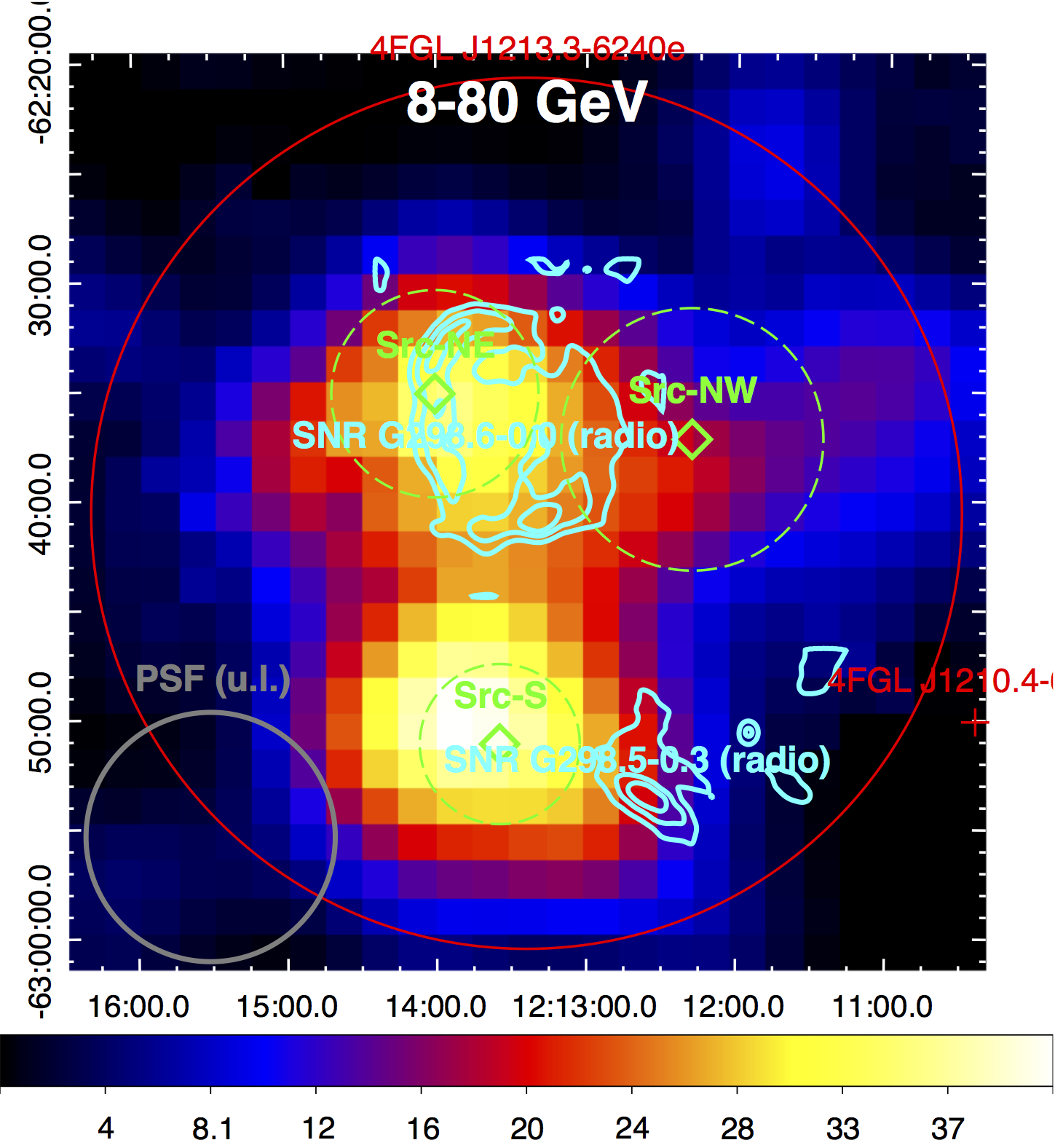}
  \includegraphics[width=55mm]{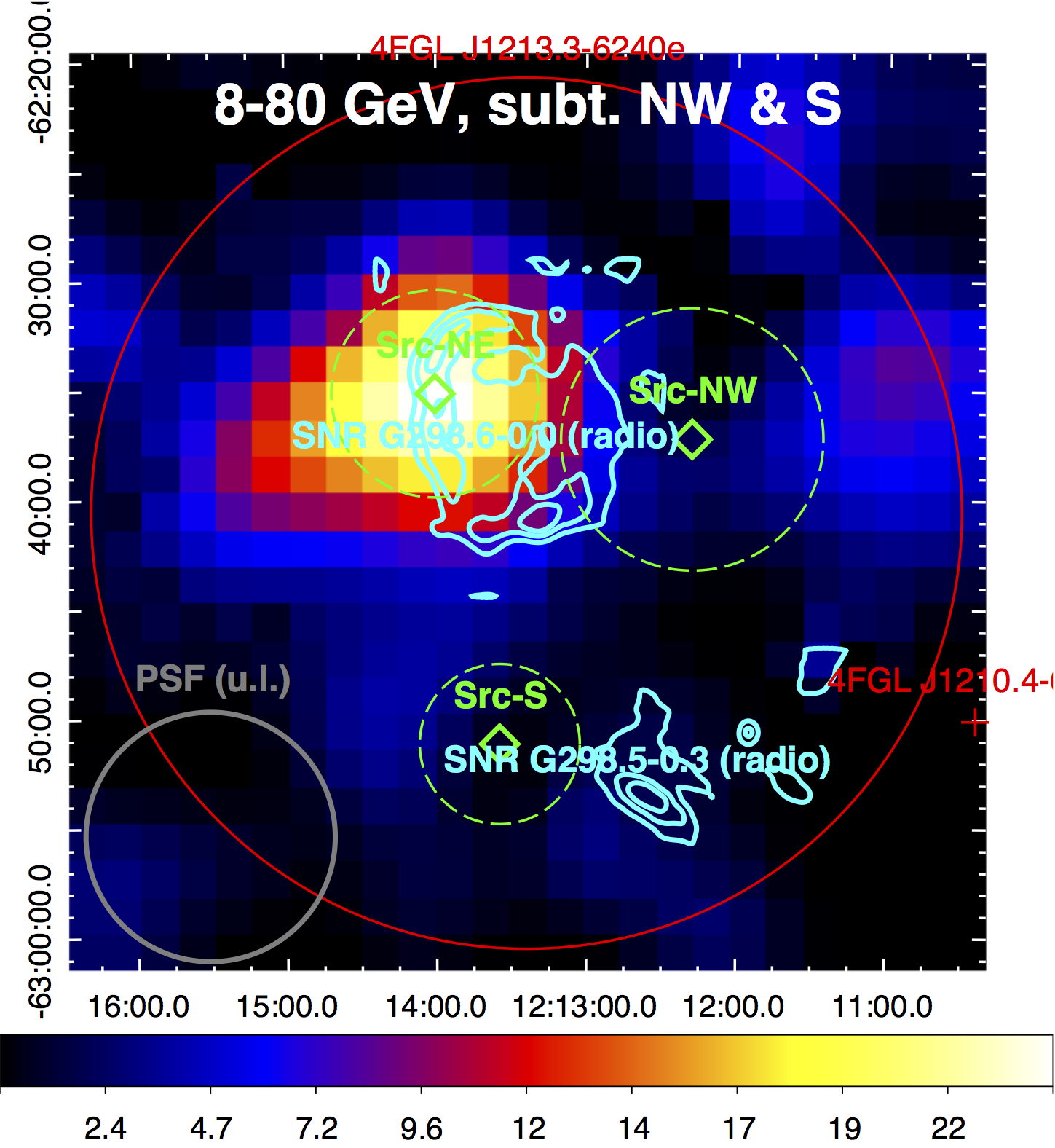}
  \includegraphics[width=84mm]{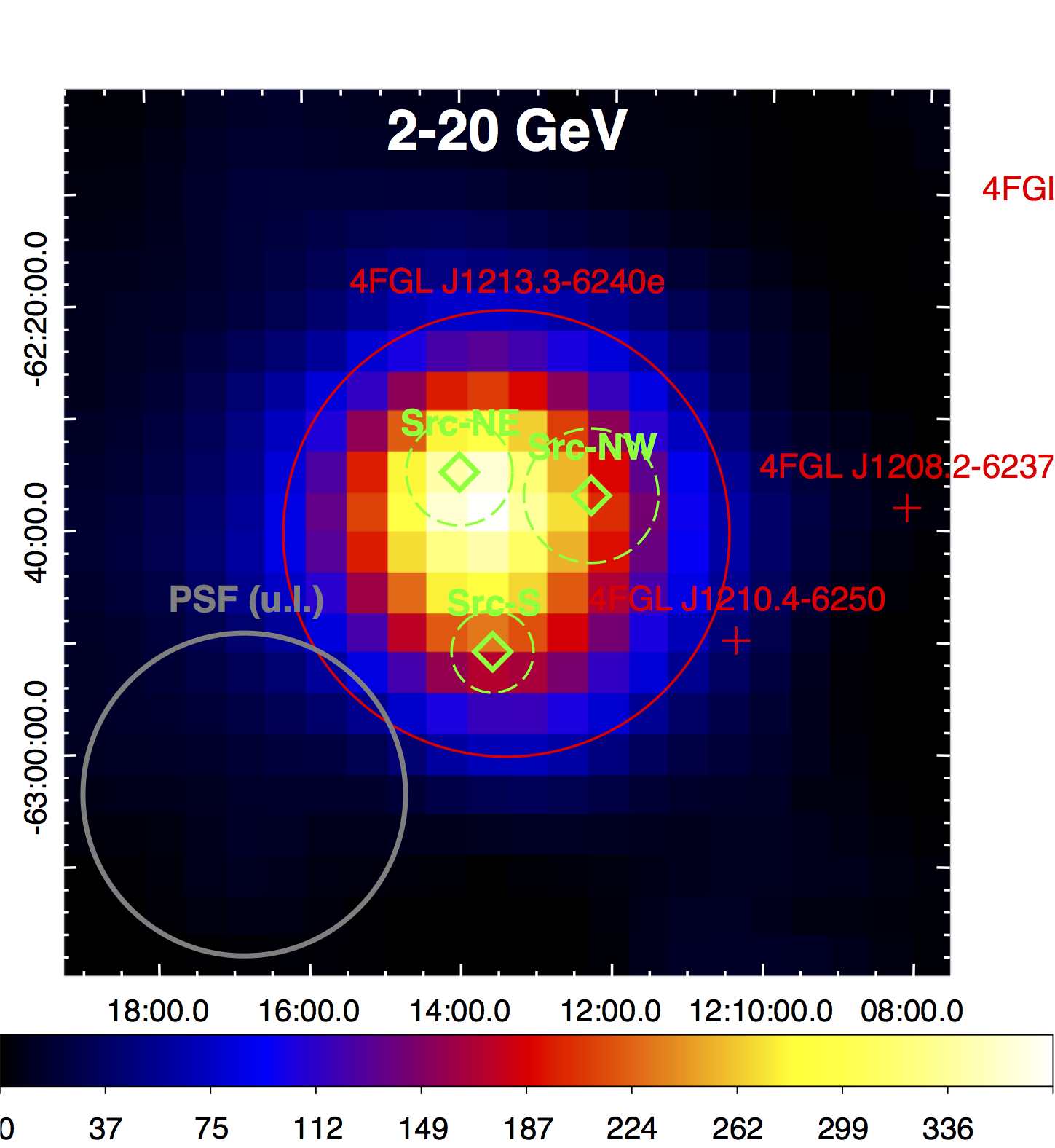}
  \includegraphics[width=84mm]{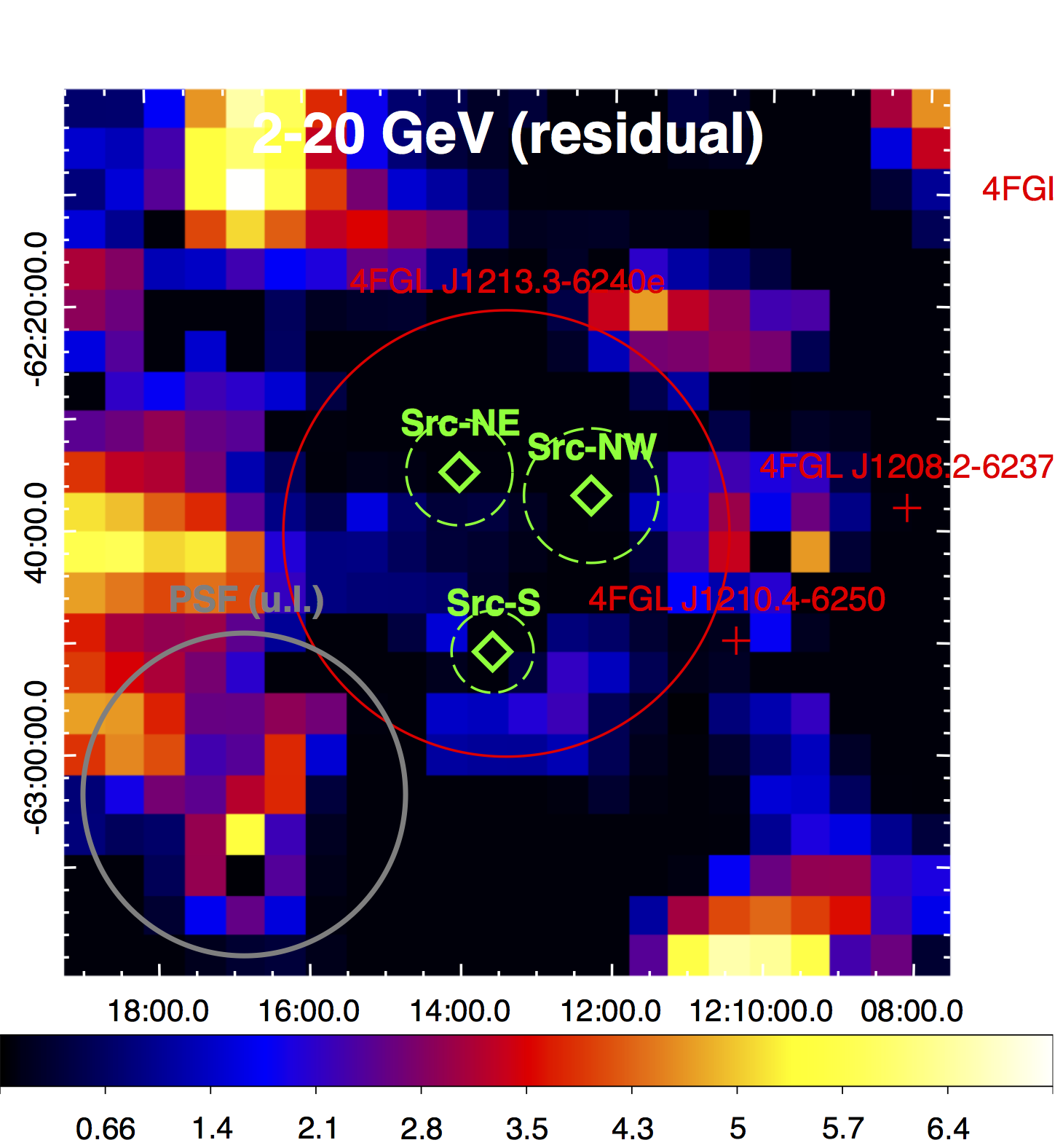}
  \end{center}
  \caption{\emph{Fermi}-LAT TS maps (for ``PSF3" data in different energy bands) of the field around G298.6$-$0.0, where all neighboring 4FGL-DR3 catalog sources and diffuse backgrounds are subtracted. The top-left panel shows a 16--300~GeV TS map, which is used to determine the positions and 95\% error circles of Src-NW and Src-S. The top-middle panel shows an 8--80~GeV TS map. The top-right panel is also created with the same 8--80~GeV dataset, but Src-NW and Src-S are  subtracted so that we determine the position and 95\% error circle of Src-NE based on this map. The 2--20~GeV map is shown on the bottom-left panel, and its corresponding residual map after subtracting Src-NE, Src-NW and Src-S is shown on the bottom-right panel. On each panel, the grey thick circle indicates the 68\% containment size of the PSF (taken from SLAC$^{\ref{slac}}$) at the applied minimum energy cut, and the positions (and dimensions) of 4FGL-DR3 catalog sources are indicated in red. The MOST 843~MHz radio contours of SNR G298.6$-$0.0 and SNR G298.5$-$0.3 \citep{Whiteoak1996} are overlaid in cyan on top panels. }
  \label{FermiMap}
\end{figure*}

In order to investigate the morphologies of $\gamma$-rays from this region, we create test-statistic (TS) maps in different energy bands (Figure~\ref{FermiMap}), where diffuse backgrounds and all 4FGL-DR3 catalog sources except 4FGL J1213.3$-$6240e are subtracted. We adopt only ``PSF3" data for the best angular resolution. On each panel, we overlay the 68\% containment circle of the PSF (see SLAC$^{\ref{slac}}$) at the applied minimum energy cut. \PKHY{Based on TS maps, we localise each centroid as a position with a peak TS value. Then, we inspect the TS distribution around each local peak, so as to  determine the error circle of each centroid at the 95\% confidence level for 4 d.o.f.\footnote{\label{error_circle}A $\chi^2$ Distribution is assumed. There are 4 d.o.f. because of the 4 variables: the right ascension, declination, flux normalization, and photon index.}, where the TS value is lower than the local peak by 9.5.} Additionally, we perform a likelihood ratio test to quantify the significance of extension for each identified sub-feature in this region. 

\PKHY{We started with analysing the $>$16~GeV data whose PSF diameter is less than one-fifth of that of 4FGL J1213.3$-$6240e.} As shown on the 16--300~GeV TS map, there are two centroids whose angular separation exceeds the PSF diameter. \PKHY{When we replace the catalogued extended source 4FGL J1213.3$-$6240e with two sources at the centroids, the maximum likelihood fit is improved with  2$\Delta$ln(likelihood)=17.1  (corresponding to $3.7\sigma$ for 2 d.o.f.).} For each spatial component, any extended morphology yields a 2$\Delta$ln(likelihood) of only $<$0.02 relative to the point-source model. Therefore, we model them as point sources, namely Src-NW (the northwest component) and Src-S (the south component), in the followings. We note that Src-NW at $(183.0752744^\circ, -62.6203894^\circ)_\mathrm{J2000}$ is only $\sim3.5^{'}$ from the western edge of our targeted SNR G298.6$-$0.0, and Src-S at $(183.394902^\circ, -62.8528466^\circ)_\mathrm{J2000}$ is only $\sim5^{'}$ from the eastern edge of the other SNR G298.5$-$0.3.

\PKHY{In order to address the bias induced by the reduced photon statistics, we also look into the $>$8~GeV data which maintains a small PSF diameter $<$30\% of that of 4FGL J1213.3$-$6240e.} From the 8--80~GeV TS map, we found an additional northeast component, namely Src-NE, which is spatially coincident with the eastern radio shell of SNR G298.6$-$0.0. Then, we re-create this map with subtracting Src-NW and Src-S, so as to localise the centroid of Src-NE, $(183.5018108^\circ, -62.5858333^\circ)_\mathrm{J2000}$, \PKHY{which remains a high detection significance of $\sim4.9\sigma$. When we replace 4FGL J1213.3$-$6240e with Src-NE, Src-NW and Src-S, the maximum likelihood fit is improved with  2$\Delta$ln(likelihood)=12.9  (corresponding to $2.52\sigma$ for 4 d.o.f.).} We also model Src-NE as a point source in the followings, since 2$\Delta$ln(likelihood)$<$0.2 for any extended morphology. 

On the 2--20~GeV TS map, because of the large PSF size, none of the aforementioned sub-features could be resolved. Nevertheless, after subtracting the three point sources Src-NE, Src-NW and Src-S, almost no residual emission is left on the whole map (the maximum TS value is only $\sim$7.2). \PKHY{Additionally, we make a comparison between the observed and modelled count maps, with a bin size of 0.1$^\circ$ that is large enough to reconcile the noise fluctuation. Among the 52 bins which are inside 4FGL J1213.3$-$6240e or overlapping its edge, only three show the observed counts less than the model predictions by 2--2.5$\sigma$, and the others show the observed and predicted counts consistent within the tolerance of $2\sigma$ uncertainties. What is more, the total observed count and total predicted count of these 52 bins are 746 and 744.9 respectively, whose difference is much less than the Poissonian statistical uncertainty. Hence, the over-/under-subtraction is less of an issue in 2--20~GeV} and it is further justifiable to replace 4FGL J1213.3$-$6240e with Src-NE, Src-NW and Src-S in subsequent analyses. 

\subsubsection{$\gamma$-ray spectroscopy}

\begin{figure*}
  \begin{center}
  \includegraphics[width=170mm]{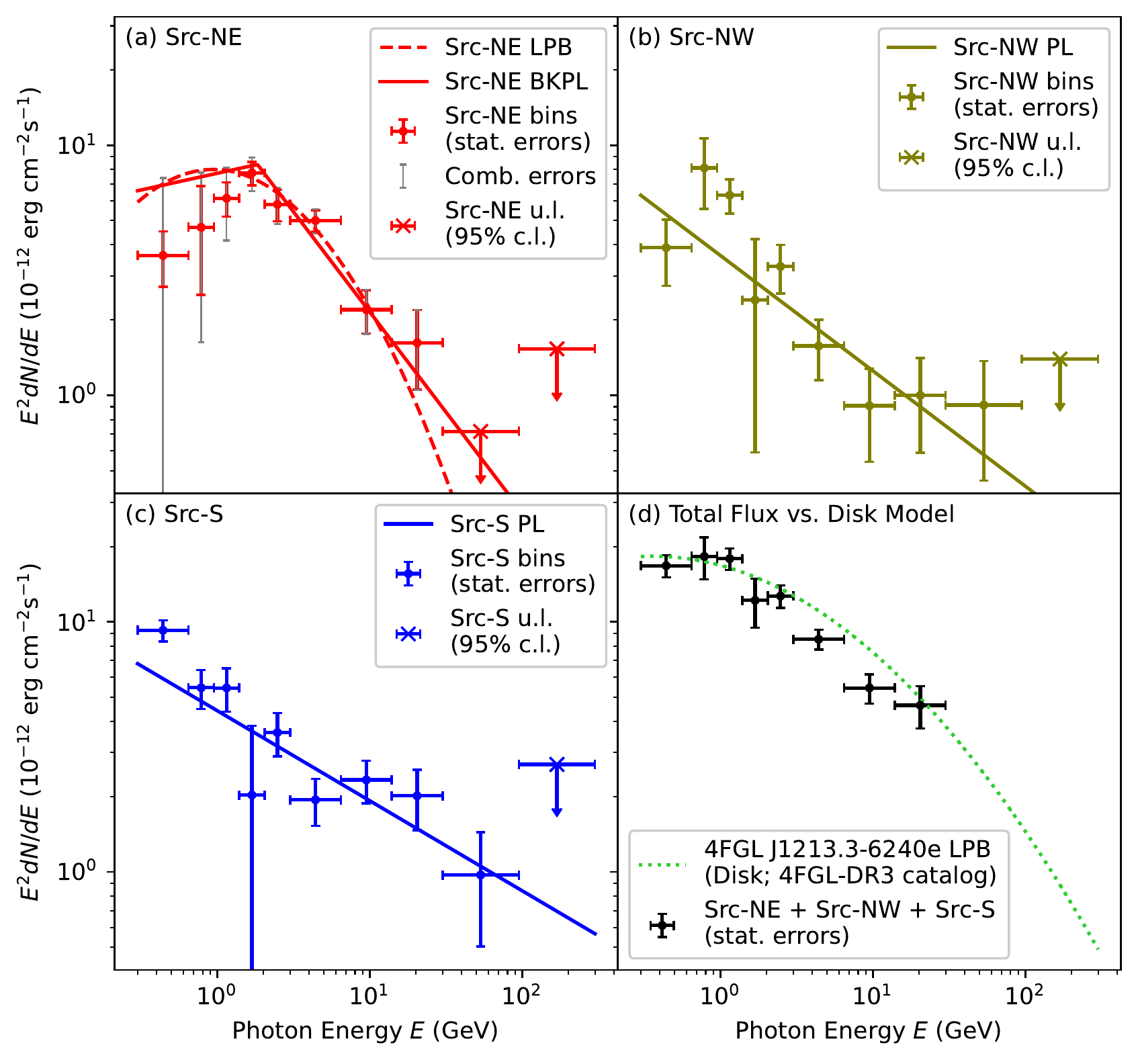}
  \end{center}
  \caption{\emph{Fermi}-LAT spectral energy distributions of (a) Src-NE,  (b) Src-NW, (c) Src-S, and (d) their total emission in comparison with the  4FGL-DR3 catalog model of 4FGL J1213.3$-$6240e. Upper limits at the 95\% confidence level are calculated for bins with TS $<$ 9. \PKHY{The combined uncertainty of each flux bin in panel (a) is defined as the statistical and systematic uncertainties added in quadrature. }}
  \label{SEDplot}
\end{figure*}

We adopt ``FRONT+BACK" data \PKHY{(i.e. photons converted into pairs in the front and back sections of the LAT tracker respectively)} for investigating the $\gamma$-ray spectra of Src-NE, Src-NW and Src-S. We enable the energy dispersion corrections for reducing the systematic effects on the spectral shape parameters. For the 0.3--300~GeV band, we perform spectral fittings with a power-law (PL) model\footnote{$N_0$ and $\Gamma$ represent the normalisation and photon index respectively.}:
\begin{equation}
\frac{dN}{dE} = N_0 \left(\frac{E}{3~\mathrm{GeV}}\right)^{-\Gamma},
\end{equation}
a log-parabola (LPB)\footnote{$\alpha$ and $\beta$ represent the photon index at 3~GeV and curvature index respectively.}:
\begin{equation}
\frac{dN}{dE} = N_0 \left(\frac{E}{3~\mathrm{GeV}}\right)^{-[\alpha+\beta \mathrm{ln}(E/3~\mathrm{GeV})]},
\end{equation}
and a broken-power-law (BKPL) model\footnote{$E_{br}$ represents the energy of the spectral break. $\Gamma_1$ and $\Gamma_2$ represent the photon indices below and above $E_{br}$ respectively. }:
\begin{equation}
\frac{dN}{dE} = N_\mathrm{0} \times \left\{
\begin{array}{ll}
	\left(\frac{E}{E_{br}}\right)^{-\Gamma_1}\,\mathrm{if}\,E<E_{br}  \\
	\left(\frac{E}{E_{br}}\right)^{-\Gamma_2}\,\mathrm{if}\,E\geq E_{br} 
\end{array}
\right. .
\end{equation}
The 0.3--300~GeV spectral properties are tabulated in Table~\ref{FermiSpecPara}, and the spectral energy distributions (including binned spectra and fitted models) are demonstrated in Figure~\ref{SEDplot}.

\begin{table*}
\tbl{$\gamma$-ray  spectral parameters  of different spatial components, determined with the \emph{Fermi}-LAT data  in 0.3--300~GeV.\footnotemark[$*$]}{
\begin{tabular}{lccc}
\hline\noalign{\vskip3pt} 
                                  & Src-NE    & Src-NW     & Src-S \\
\hline\noalign{\vskip3pt} 
 \multicolumn{4}{c}{\underline{Power-Law (PL)}}                       \\
$\Gamma$    & 2.364 $\pm$ 0.036    & 2.461  $\pm$ 0.071    & 2.360  $\pm$ 0.054    \\
Flux\footnotemark[$\dagger$] (10$^{-9}$~ph~cm$^{-2}$~s$^{-1}$) & 16.21 $\pm$ 0.85     & 8.99   $\pm$ 0.90     & 10.41  $\pm$ 0.90     \\
TS  & 690.3            & 177.4            & 313.7            \\
\hline\noalign{\vskip3pt} 
 \multicolumn{4}{c}{\underline{Log-Parabola (LPB)}}                   \\
$\alpha$    & 2.534 $\pm$ 0.070    ($^{+0.108        }_{-0.092   }$)    & 2.458  $\pm$ 0.072    & 2.355  $\pm$ 0.050    \\
$\beta$    & 0.230 $\pm$ 0.045    ($^{+0.084        }_{-0.046   }$)    & -0.007 $\pm$ 0.039    & -0.038 $\pm$ 0.028    \\
Flux (10$^{-9}$~ph~cm$^{-2}$~s$^{-1}$) & 14.08 $\pm$ 0.93     ($^{+7.38         }_{-4.14   }$)     & 9.02   $\pm$ 0.93     & 10.82  $\pm$ 0.93     \\
TS  & 730.3            & 177.4            & 315.3            \\
\hline\noalign{\vskip3pt} 
 \multicolumn{4}{c}{\underline{Broken-Power-Law (BKPL)}}              \\
$\Gamma_1$   & 1.866 $\pm$ 0.135    ($^{+0.260        }_{-0.353   }$)    & -1.129  $\pm$ 7.174    & 2.463  $\pm$ 0.113    \\
$E_{br}$ (MeV) & 1832  $\pm$ 440      ($^{+199          }_{-94    }$)      & 400  $\pm$ 178  & 4003   $\pm$ 1480     \\
$\Gamma_2$   & 2.801 $\pm$ 0.132    ($^{+0.078        }_{-0.029   }$)    & 2.512  $\pm$ 0.079    & 2.177  $\pm$ 0.124    \\
Flux (10$^{-9}$~ph~cm$^{-2}$~s$^{-1}$) & 14.09 $\pm$ 0.95     ($^{+7.49         }_{-4.24   }$)     & 8.50   $\pm$ 2.18     & 10.90  $\pm$ 0.97     \\
TS  & 731.4            & 179.0            & 316.2            \\
\hline\noalign{\vskip3pt} 
\end{tabular}

}\label{FermiSpecPara}
  \begin{tabnote}
 \footnotemark[$*$] The uncertainties without parentheses are statistical ($1\sigma$), and those inside parentheses are systematic.   \\
 \footnotemark[$\dagger$] The fluxes are integrated from 0.3~GeV to 300~GeV.
  \end{tabnote}
\end{table*}

For the Src-NE spectrum, both LPB and BKPL models are preferred over PL by $\ge6.3\sigma$. The TS values yielded by LPB and BKPL differ by only $\sim$1.1, making it unclear whether the spectral turnover is a gradual curvature or a sharp peak. \PKHY{The apparent discrepancy between the flux bins and BKPL model below the spectral break is discussed in \S\ref{caveat}.} Both Src-NW and Src-S spectra are satisfactorily described by PL, so that the extra parameters of LPB and BKPL are not significantly required (only $\le1.6\sigma$). 

In order to cross-check the significance of the spectral curvature/break for Src-NE, we repeat the LPB and BKPL fittings for Src-NE with the following adjustments: (i) shifting the normalisation of the Galactic diffuse background by $\pm$5\%; (ii) discarding ``PSF0" data (i.e. data of poorest angular resolution); (iii) discarding ``EDISP0" data (i.e. data of poorest energy resolution). Each of these adjustments is implemented separately. We record the difference in each parameter value for each adjustment. We quantify the systematic uncertainty of each parameter (appended in Table~\ref{FermiSpecPara}) as the differences added in quadrature. \PKHY{We follow a similar scheme to determine the background-associated systematic uncertainties for the flux bins of Src-NE as well, in an attempt to account for the discrepancy between its binned spectrum and BKPL model (\S\ref{caveat}). }

For Src-NE, after combining the statistical and systematic uncertainties in quadrature, the curvature index $\beta$ of LPB is still greater than 0 by $\sim3.6\sigma$, the break energy $E_{br}$ of BKPL is still higher than 0.3~GeV (the applied minimum energy cut) by $\sim3.4\sigma$, and the photon index $\Gamma_2$ above the break is still softer than $\Gamma_1$ below the break by $\sim2.9\sigma$. We hereby confirm that the spectral turnover of Src-NE  is genuine. The differential flux of Src-NE peaks at $E_{br}=1.83\pm0.44_{stat}{^{+0.20}_{-0.09}}_{sys}$~GeV. Based on the binned spectra, we constrain the peak energy of the differential flux to be $\lesssim$1.2~GeV and $\lesssim$0.5~GeV for Src-NW and Src-S respectively. 

We repeat the PL and LPB fittings for the three spatial components with 2--300~GeV data, so that the data below the spectral turnover of Src-NE is excluded. The results are tabulated in Table~\ref{2GeVSpecPara}. It turns out that PL is sufficient to describe the 2--300~GeV spectrum of each component, while LPB only improves the fits by $\le1.1\sigma$. Noticeably, Src-NE has the softest spectrum while Src-S has the hardest spectrum. 

\begin{table*}
\tbl{$\gamma$-ray  spectral parameters  of different spatial components, determined with the \emph{Fermi}-LAT data  in 2--300~GeV.\footnotemark[$*$]}{
\begin{tabular}{lccc}
\hline\noalign{\vskip3pt} 
                                  & Src-NE    & Src-NW     & Src-S \\
\hline\noalign{\vskip3pt} 
 \multicolumn{4}{c}{\underline{Power-Law (PL)}}                       \\
$\Gamma$    & 2.814 $\pm$ 0.110    ($^{+0.146         }_{-0.016    }$)    & 2.462  $\pm$ 0.161    ($^{+0.156         }_{-0.010    }$)    & 2.252  $\pm$ 0.109    ($^{+0.049         }_{-0.034    }$)    \\
Flux\footnotemark[$\dagger$] (10$^{-9}$~ph~cm$^{-2}$~s$^{-1}$) & 1.36  $\pm$ 0.09     ($^{+0.23          }_{-0.06    }$)     & 0.55   $\pm$ 0.08     ($^{+0.06          }_{-0.01    }$)     & 0.73   $\pm$ 0.08     ($^{+0.08          }_{-0.05    }$)     \\
TS  & 344.1              & 75.7                & 161.0               \\
\hline\noalign{\vskip3pt}
 \multicolumn{4}{c}{\underline{Log-Parabola (LPB)}}                   \\
$\alpha$    & 2.742 $\pm$ 0.212      & 2.788  $\pm$ 0.175      & 2.341  $\pm$ 0.138      \\
$\beta$    & 0.050 $\pm$ 0.130      & -0.122 $\pm$ 0.056      & -0.033 $\pm$ 0.046      \\
Flux (10$^{-9}$~ph~cm$^{-2}$~s$^{-1}$) & 1.36  $\pm$ 0.10       & 0.57   $\pm$ 0.08       & 0.74   $\pm$ 0.08       \\
TS  & 344.2              & 76.9                & 161.1              \\
\hline\noalign{\vskip3pt}
\end{tabular}

}\label{2GeVSpecPara}
  \begin{tabnote}
 \footnotemark[$*$] The uncertainties without parentheses are statistical ($1\sigma$), and those inside parentheses are systematic.   \\
 \footnotemark[$\dagger$] The fluxes are integrated from 2~GeV to 300~GeV.
  \end{tabnote}
\end{table*}

We follow the aforementioned scheme to determine the systematic uncertainties of PL parameters for the three components (appended in Table~\ref{2GeVSpecPara}). After combining the statistical and systematic uncertainties in quadrature, the 2--300~GeV photon index $\Gamma$ of Src-NE is still softer than that of Src-S by $\sim3.4\sigma$. We also note that the  2--300~GeV  integrated flux of Src-NE is higher than that of Src-NW by $\sim5.5\sigma$, considering their combined uncertainties. 

We compute the binned spectrum of the total emission of Src-NE, Src-NW and Src-S and compare it with the 4FGL-DR3 catalog model of 4FGL J1213.3$-$6240e (see Figure~\ref{SEDplot}(d)), which assumes a LPB spectral shape and a disk spatial morphology. It turns out that they are essentially consistent with each other, despite the maximum deviation of $\sim30\%$ at $\sim$9.5~GeV. 

\subsection{Chandra-ACIS results}\label{Xrayresults}

\subsubsection{X-ray spatial morphology}

In X-ray spatial analyses, considering the Suzaku result of \citet{Bamba2016}, we adopt the 1--4~keV data so as to roughly maximise the signal-to-noise ratio of the diffuse emission at/around SNR G298.6$-$0.0. \PKHY{In a search for point sources by} the CIAO tool ``wavdetect", we identified five point-like features (Table~\ref{XrayBkgPt}) in the whole field of view. Each of their apparent source ellipses indicates the combination of the PSF size and positional uncertainty \PKHY{in centroid localisation, while their actual extensions are negligibly small compared with the PSF}.

\begin{table*}
\tbl{Spatial (and spectral) properties of five point-like X-ray sources identified with ``wavdetect". }{
    \begin{tabular}{lccccc}
    \hline\noalign{\vskip3pt} 
          & Pt1 & Pt2    & Pt3 & Pt4    & Pt5    \\ \hline
RA ($^\circ$)     &     183.743     &     183.695     &     183.457     &     183.331     &     183.208    \\ 
Dec ($^\circ$)     &     -62.598     &     -62.552     &     -62.667     &     -62.635     &     -62.513    \\ 
        $R_\mathrm{apx}$\footnotemark[$a$] ($^{''}$) & 40.9 & 8.7    & 19.1 & 7.7    & 32.3    \\ \hline
           \multicolumn{6}{c}{\underline{Source Spectrum}}    \\ 
        $N_\mathrm{H,src}$\footnotemark[$b$] (10$^{22}$~cm$^{-2}$) & -- & 0.22 $\pm$ 0.12 & -- & 0.35 $\pm$ 0.17 & --    \\ 
        $\Gamma_\mathrm{src}$\footnotemark[$c$] & -- & 2.13 $\pm$ 0.26 & -- & 3.45 $\pm$ 0.57 & --    \\ 
        $F_\mathrm{src}$\footnotemark[$d$] (10$^{-15}$~erg~cm$^{-2}$~s$^{-1}$) & -- & 96 $\pm$ 14 & -- & 14 $\pm$ 4 & --    \\ \hline
           \multicolumn{6}{c}{\underline{Background Spectrum}}    \\ 
        $N_\mathrm{H,bkg}$\footnotemark[$e$] (10$^{22}$~cm$^{-2}$) & -- & 0 $\pm$ 0.06 & -- & 0 $\pm$ 0.12 & --    \\ 
        $\Gamma_\mathrm{bkg}$\footnotemark[$f$] & -- & 0.88 $\pm$ 0.30 & -- & 1.08 $\pm$ 0.37 & --    \\ 
        $F_\mathrm{bkg}$\footnotemark[$g$] (10$^{-15}$~erg~cm$^{-2}$~s$^{-1}$) & -- & 45 $\pm$ 14 & -- & 26 $\pm$ 10 & --    \\ 
\hline\noalign{\vskip3pt} 
    \end{tabular}
}\label{XrayBkgPt}
  \begin{tabnote}
 \footnotemark[$a$] Approximate radius of an apparent source ellipse, defined as the square-root of the product of the semi-major and semi-minor axes  determined by ``wavdetect"  \PKHY{(i.e. we approximate an ellipse as a circle with the same area)}. This quantifies the combination of the PSF size and positional uncertainty \PKHY{in centroid localisation}.   \\
 \footnotemark[$b$] Equivalent hydrogen column density for photoelectric absorption at the direction of the source component.  \\
 \footnotemark[$c$] Power-law index of the unabsorbed source spectrum.  \\
 \footnotemark[$d$] Unabsorbed 2--10~keV source flux predicted by the power-law.  \\
 \footnotemark[$e$] Equivalent hydrogen column density for photoelectric absorption at the direction of the background component.  \\
 \footnotemark[$f$] Power-law index of the unabsorbed background spectrum.  \\
 \footnotemark[$g$] Unabsorbed 2--10~keV background flux predicted by the power-law.  \\
  \end{tabnote}
\end{table*}

\PKHY{These source ellipses are all beyond the 95\% error circles of our identified $\gamma$-ray sources Src-NE, Src-NW and Src-S, making it difficult to relate these point-like X-ray sources to those GeV $\gamma$-ray emissions (a more detailed discussion is in \S\ref{DiscSumm}).} Among the five point-like features, only Pt2 and Pt4 have their source ellipses smaller than or comparable to the wing part of the Chandra-ACIS PSF\footnote{Understanding the Chandra PSF:
\url{https://cxc.cfa.harvard.edu/ciao/PSFs/psf_central.html}} (a radius of $\sim10^{''}$), \PKHY{because  their higher photon statistics lead to preciser localisations of their centroids}. The source ellipses of Pt1, Pt3 and Pt5 have their approximate radii greater than or about a double of the on-axis PSF wing's, reflecting their large positional uncertainties due to their off-axis positions and the lack of photon statistics.

\begin{figure}
  \begin{center}
  \includegraphics[width=80mm]{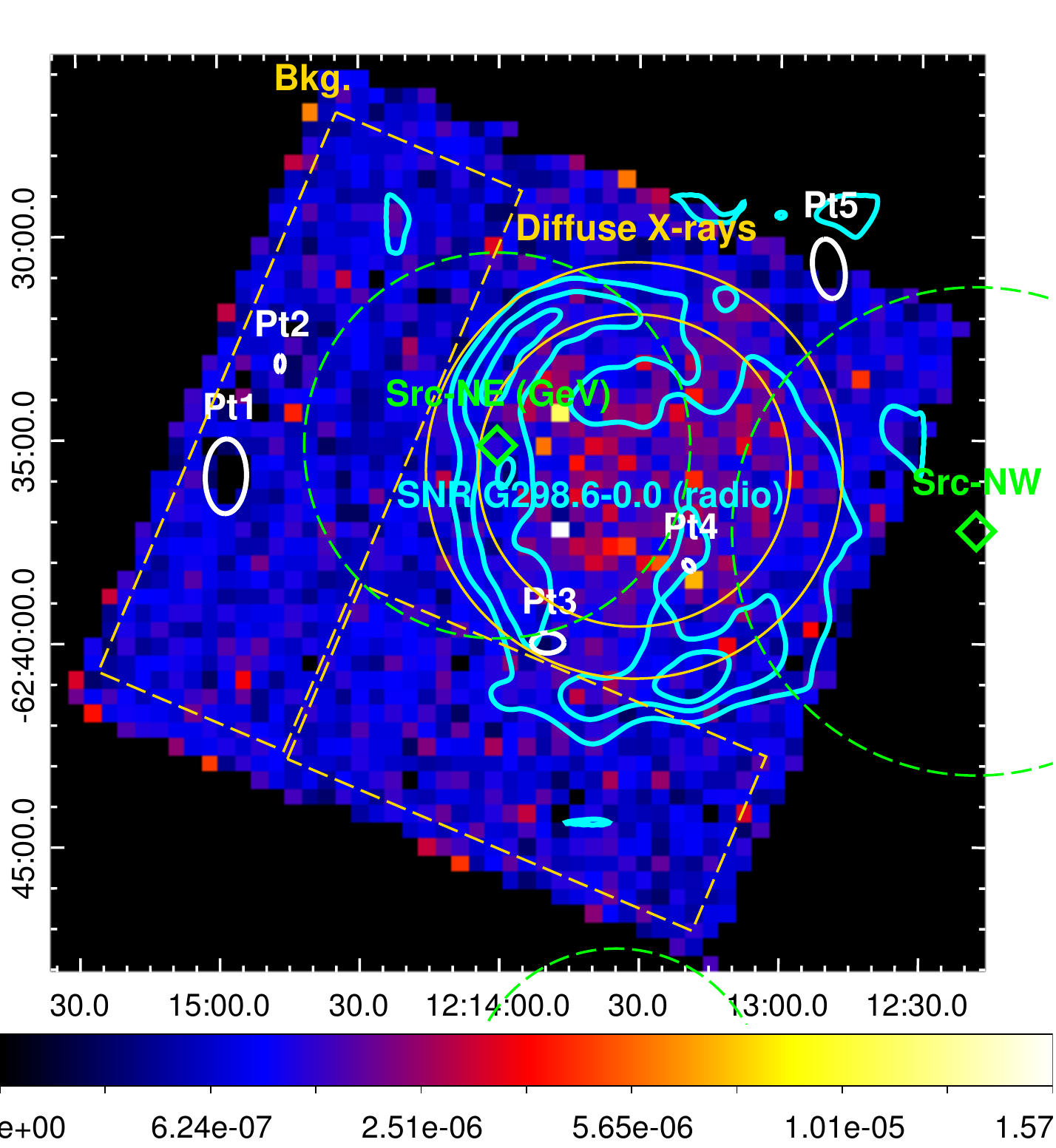}\\
  \includegraphics[width=80mm]{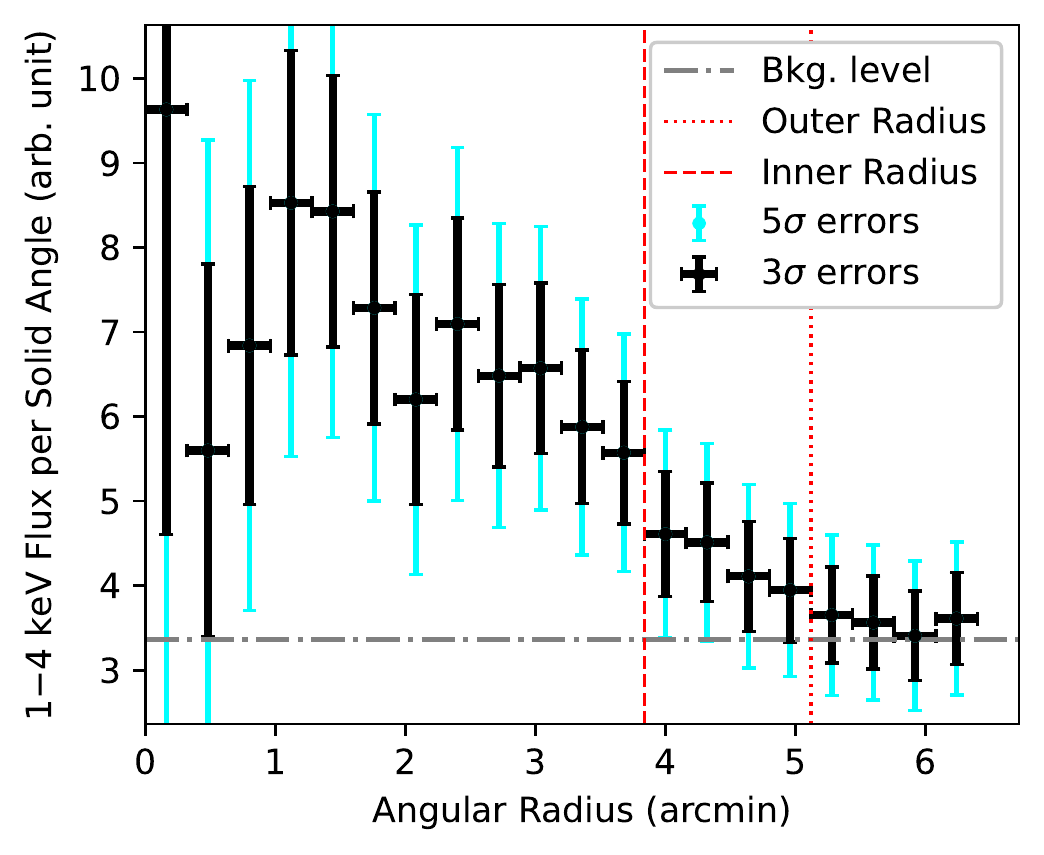}
  \end{center}
  \caption{(top) Chandra-ACIS flux map (i.e. exposure-corrected map) of the field around G298.6$-$0.0, in 1--4~keV. It is overlaid with the MOST 843~MHz radio contours of G298.6$-$0.0 \citep{Whiteoak1996} in cyan. The  centroids and 95\% error circles of our identified $\gamma$-ray sources  are indicated in green. The color-bar represents the flux in the unit of ph~cm$^{-2}$~s$^{-1}$ and is in a square-root (``sqrt") scale. The two gold concentric circles indicate the inner and outer regions of the diffuse X-ray source, which are defined according to the bottom panel. The combination of the two gold dashed boxes is our chosen background region. Five point-like background sources which have been subtracted are indicated as white thick ellipses determined by the ``wavdetect" algorithm. (bottom) Radial brightness profile around the 1--4~keV centroid. The $3\sigma$ uncertainties (black) and the background flux (the grey horizontal line) are used to determine the outer radius (the red dotted line), while the $5\sigma$ uncertainties (cyan) and the background flux are used to determine the inner radius (the red dashed line). }
  \label{XrayMap}
\end{figure}

Then, we create a flux map (i.e. an exposure-corrected map; the top panel of Figure~\ref{XrayMap}) with excluding the data in the determined elliptical regions of these point-like features. We use ds9 to determine the flux-weighted centroid of this map, and we apply the CIAO command ``dmextract" on the event file to compute a radial brightness profile around this centroid (the bottom panel of Figure~\ref{XrayMap}). 

\PKHY{As the separation from the centroid increases}, we notice a gradually decreasing trend of the brightness. We define the ``inner" and ``outer" regions of the diffuse source \PKHY{(i.e. the two gold concentric circles in Figure~\ref{XrayMap})} in this way: beyond the inner region, the brightness starts to be consistent with the background level within the tolerance of $\sim5\sigma$ uncertainties; beyond the outer region, the brightness is no longer higher than the background level by $\gtrsim3\sigma$. The dimension of this diffuse X-ray source and the radio-continuum region of SNR G298.6$-$0.0 largely overlap each other, as shown in the top panel of Figure~\ref{XrayMap}. 

For the first time, we notice that the centroid position of this extended keV source is slightly tilted to the northeast part of the SNR. It shows no significant emission ($<3\sigma$) at the south, southwest and west edges of the SNR. Such an X-ray morphology is a stark contrast to the radio-continuum morphology \citep{Whiteoak1996} whose southwest shell is the brightest region.

\subsubsection{X-ray spectroscopy}

\begin{figure}
  \begin{center}
  \includegraphics[width=80mm]{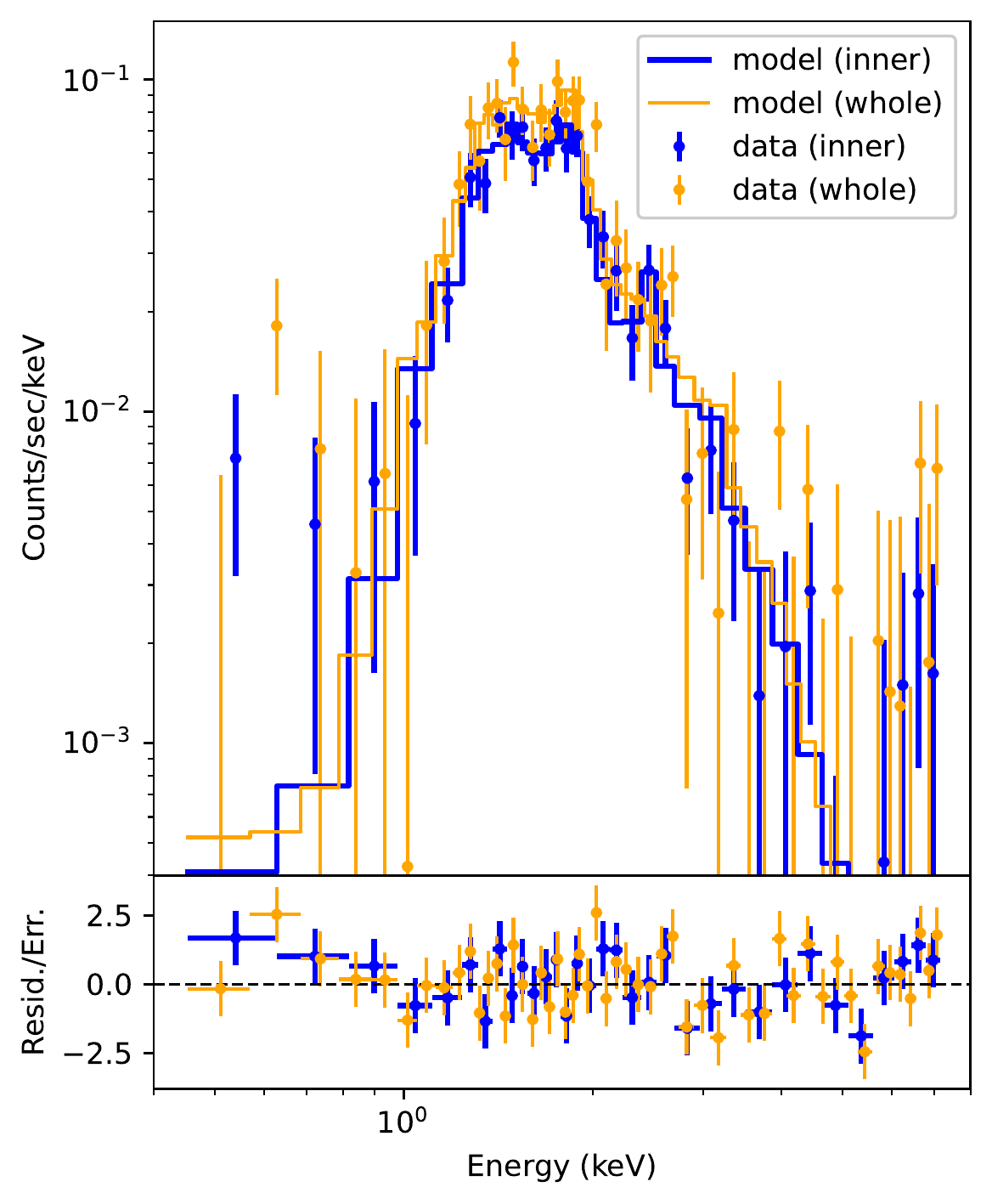}
  \end{center}
  \caption{(top) Background-subtracted Chandra-ACIS spectra  of the diffuse X-ray source associated with G298.6$-$0.0, and the best-fit $\mathrm{VAPEC{\times}PHABS}$ models. The data and model for our defined ``inner" region are plotted in blue, while those for our defined ``whole" (inner+outer) region are plotted in orange. (bottom) Residual (data$-$model) divided by the corresponding uncertainty.}
  \label{XraySpec}
\end{figure}

We look into the 0.5--7~keV spectra of our defined ``inner"  and ``whole" (inner+outer) regions, where the background emissions from the source-free region (see Figure~\ref{XrayMap}) are subtracted. We fit the $\mathrm{VAPEC{\times}PHABS}$ model (i.e. an emission spectrum from plasma in collisional ionisation equilibrium, with interstellar absorption) to each spectral dataset. We have seven free parameters in the fittings: the equivalent hydrogen column density ($N_\mathrm{H}$) for photoelectric absorption, the plasma temperature ($kT$), the emission measure (EM; i.e. the normalisation), the abundances of magnesium (Mg), silicon (Si), sulphur (S) and iron (Fe). The data points and model lines are plotted in Figure~\ref{XraySpec}, and the parameter values are tabulated in Table~\ref{xspec_fit}. 

\begin{table}
\tbl{Spectral fittings for the Chandra-ACIS data of the diffuse X-ray source in 0.5--7~keV, with the $\mathrm{VAPEC{\times}PHABS}$ model.\footnotemark[$\dagger$] }{
\begin{tabular}{lcc}
\hline\noalign{\vskip3pt}
                                 & Inner Region       & Whole Region       \\
\hline\noalign{\vskip3pt}
$N_\mathrm{H}$ (10$^{22}$~cm$^{-2}$) & 1.93     $\pm$     0.27 & 2.25     $\pm$     0.55 \\
$kT$ (keV)                         & 0.71     $\pm$     0.10 & 0.64     $\pm$     0.11 \\
Mg                               & 0.45     $\pm$     0.25 & 0.48     $\pm$     0.32 \\
Si                               & 0.27     $\pm$     0.11 & 0.26     $\pm$     0.10 \\
S                                & 0.41     $\pm$     0.22 & 0.02     $\pm$     0.17 \\
Fe                               & 0        $\pm$     0.04 & 0.18     $\pm$     0.24 \\
EM (cm$^{-5}$)\footnotemark[$*$] & 8.9      $\pm$     3.8  & 15.8     $\pm$     8.7  \\
$\chi^2/d.o.f.$                      & 30.1         / 26   & 63.3         / 45  \\
\hline\noalign{\vskip3pt} 
\end{tabular}
}\label{xspec_fit}
  \begin{tabnote}
 \footnotemark[$\dagger$] The abundances of elements are in solar units. \\
 \footnotemark[$*$] Emission measure, defined as $10^{-11}(4\pi D^2)^{-1}\int n_en_{\rm H} dV$. 
  \end{tabnote}
\end{table}

We compare our spectral fitting results for the two defined regions and that  of \citet{Bamba2016} for Suzaku data altogether. For all parameters except S, three sets of solutions are mutually consistent with each other within the tolerance of $1\sigma$ uncertainties. The S value we obtain for the whole region appears to be smaller than that for the inner region as well as that obtained by \citet{Bamba2016}, but the significance of the difference is only $<2\sigma$. 

We also look into the 0.5--7~keV spectra of our identified point-like features, whose spectral files are also generated in the ``wavdetect" procedure. Among these five sources, only two (respectively labelled as Pt2 and Pt4 in Figure~\ref{XrayMap}) have sufficient photon statistics for spectral fittings. In view of the relatively low signal-to-noise ratios, we adopt ``cstat" as the statistic function, we perform fittings on ungrouped data, and we assume an absorbed power-law model for the Pt2 and Pt4 spectra as well as the corresponding background spectra. 
The major outcome of these fittings (appended in Table~\ref{XrayBkgPt}) is that all $N_\mathrm{H}$ values are much smaller (at $>4.5\sigma$ significances) than the $N_\mathrm{H}$ we obtained for the diffuse source's inner region and the $N_\mathrm{H}$ reported by \citet{Bamba2016} on the diffuse source.

\subsection{NANTEN CO and ATCA-Parkes HI results}\label{radioresults}

\begin{figure}
  \begin{center}
  \includegraphics[width=80mm]{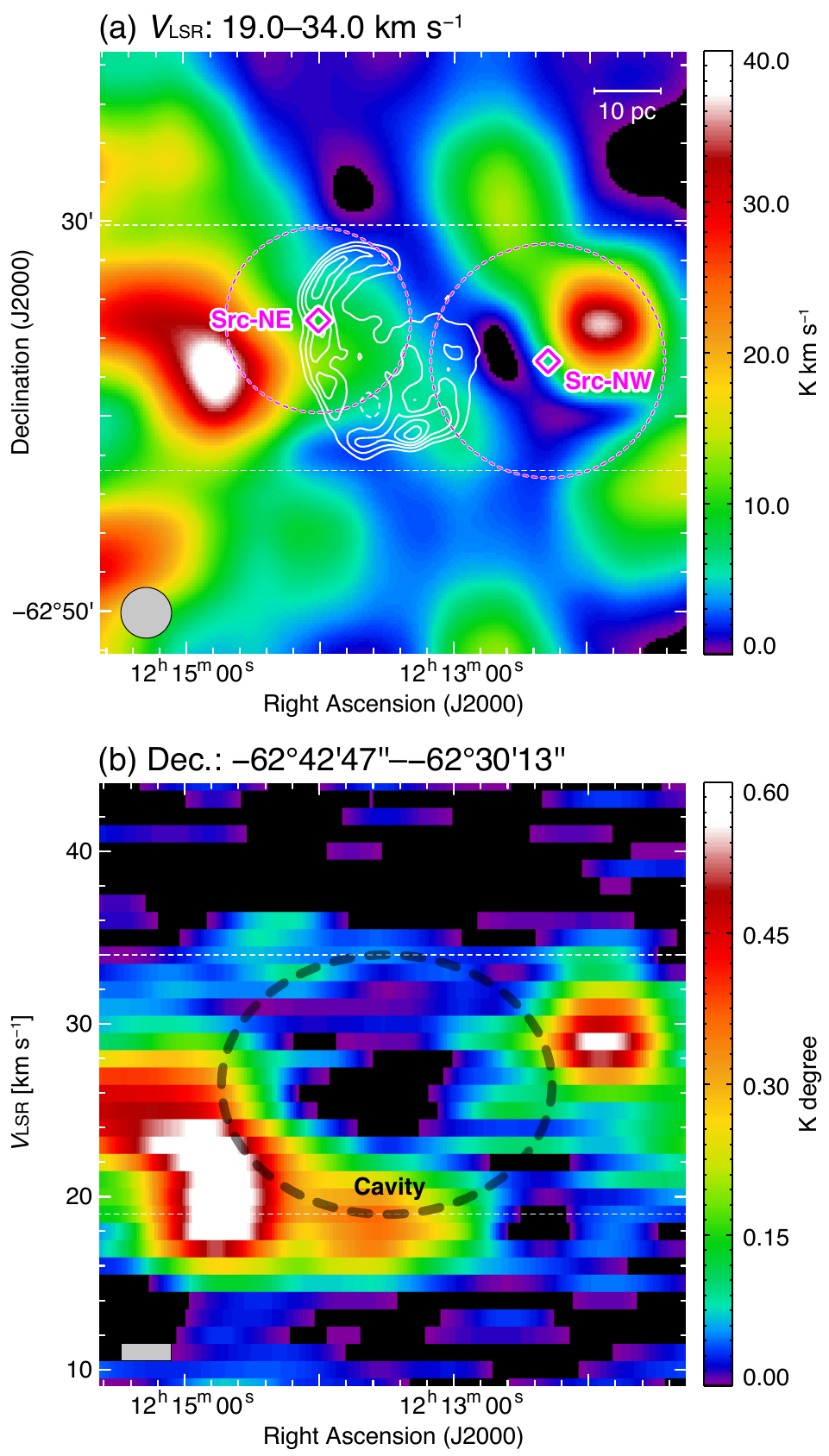}
  \end{center}
  \caption{(a) Integrated-intensity map of the $^{12}$CO($J$=1--0) emission line, observed with NANTEN, towards SNR G298.6$-$0.0. The $V_\mathrm{LSR}$ range for integration is from 19.0~km~s$^{-1}$ to 34.0~km~s$^{-1}$, which is most likely associated with the SNR. The superimposed white contours indicate the MOST 843~MHz radio-continuum emission \citep{Whiteoak1996}. The  centroids and 95\% error circles of $\gamma$-ray sources Src-NE and Src-NW are indicated in magenta. The grey circle at the bottom-left corner of the map indicates the half-power beam width of NANTEN. (b) Position-velocity ($p-v$) diagram of the CO line. The declination range for integration is from $-62^{\circ}42^{'}47^{''}$ to $-62^{\circ}30^{'}13^{''}$. We highlight the cavity-like structure with a thick-dashed ellipse.
}
  \label{G298_CO_pv}
\end{figure}

Figure~\ref{G298_CO_pv} shows the integrated-intensity map of the NANTEN $^{12}$CO($J$=1--0) data superimposed with the radio-continuum contours and the position-velocity ($p-v$) diagram of CO towards SNR G298.6$-$0.0. \PKHY{The integrated declination range (from $-62^{\circ}42^{'}47^{''}$ to $-62^{\circ}30^{'}13^{''}$) for this $p-v$ diagram closely matches the declination range of the SNR extension.} Interestingly, this $p-v$ diagram demonstrates a cavity-like structure towards G298.6$-$0.0 at $V_\mathrm{LSR}$ of 27$\pm$8~km~s$^{-1}$. It is noteworthy that this cavity has an extent in right ascension that is roughly consistent with the right ascension range of the SNR. Besides, the integrated-intensity map of CO demonstrates  significant  peaks at around $(12^{h}11^{m}48^{s}, -62^{\circ}35^{'}30^{''})_\mathrm{J2000}$ and  $(12^{h}14^{m}48^{s}, -62^{\circ}38^{'}00^{''})_\mathrm{J2000}$, which roughly coincide with the $\gamma$-ray centroids of Src-NE and Src-NW.

\begin{figure*}
  \begin{center}
  \includegraphics[width=170mm]{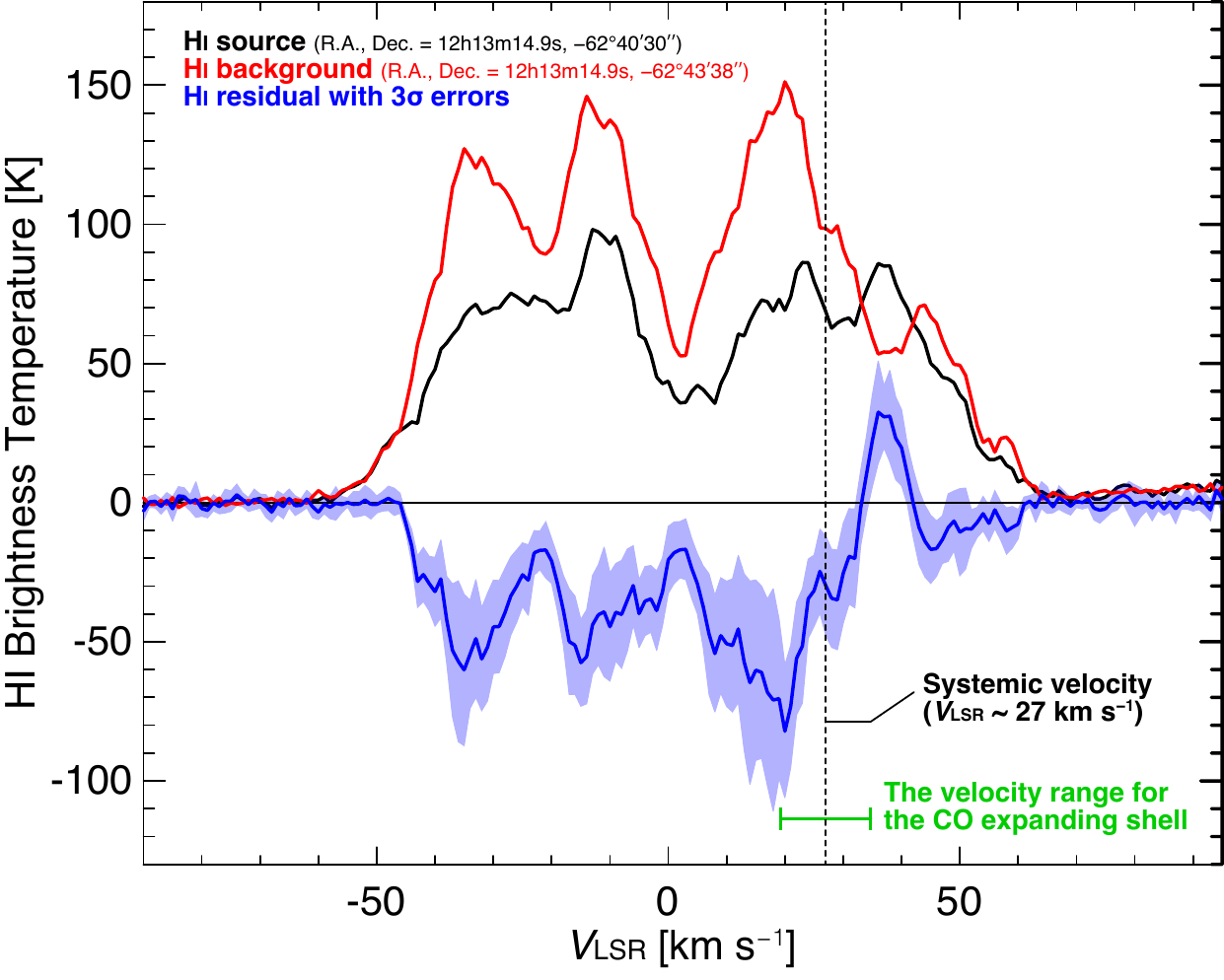}
  \end{center}
  \caption{Spectra of the HI (1.4~GHz) line, reconstructed from the combined ATCA-Parkes data, towards G298.6$-$0.0 (black) and a background region (red) respectively. The extraction region for each spectrum is approximately a square box of side $\sim100^"$. Additionally, the residual (source$-$background) is plotted as a blue line sandwiched within the blue-shaded $3\sigma$ uncertainty range.
}
  \label{G298_HI_spec}
\end{figure*}

Figure~\ref{G298_HI_spec} shows the typical HI spectra towards SNR G298.6$-$0.0 and a background region respectively. We found  a hint of an HI absorption feature in a  broad $V_\mathrm{LSR}$ range approximately from $-$46~km~s$^{-1}$ to +32~km~s$^{-1}$, where the positive-$V_\mathrm{LSR}$ end is well consistent with the CO cavity. 

We will explore the relation among the CO cavity, HI absorption and the SNR, as well as the relation between the MC clumps and the $\gamma$-ray emissions, in \S\ref{hadronic_SNR--MC}.

\section{Discussion \& Summary}\label{DiscSumm}

From the \emph{Fermi}-LAT TS maps (Figure~\ref{FermiMap}), we found that the extended GeV $\gamma$-ray source 4FGL J1213.3$-$6240e in the 4FGL-DR3 catalog is decomposable into three spatial components (Src-NE, Src-NW and Src-S) with point-like morphologies. Src-NE has a much higher spectral peak energy than Src-S, and the 2--300~GeV photon index of Src-NE is significantly softer than that of Src-S. Such differences in spectral shape suggest different origins of the responsible cosmic-rays. Src-NE is  at the eastern shell of SNR G298.6$-$0.0, and Src-NW is adjacent to the western edge of this SNR. Therefore, we will explore the SNR--MC interaction scenarios for Src-NE and Src-NW. Src-S is very close to the other SNR G298.5$-$0.3, suggesting a possible association as well, but the discussions on this pair are beyond the scope of this paper. 

The Chandra-ACIS flux map (Figure~\ref{XrayMap}) demonstrates a diffuse keV X-ray source which is highly spatially coincident with SNR G298.6$-$0.0. Our defined inner region of this X-ray feature is a brighter and filled structure inside the radio SNR shell, confirming that G298.6$-$0.0 is a mixed-morphology SNR (as put forward by \cite{Bamba2016}). 

It is worth mentioning that, in our X-ray spectral fitting results, the average sulphur abundance of  the whole diffuse X-ray source seems to be lower than that of the inner region. If this apparent difference is genuine, then it possibly implies that the ejecta keep an onion-like structure of nuclear burning, resulting in a concentration of the sulphur abundance in the inner region. Nevertheless, because of the large uncertainties, we can only claim a low significance of $<2\sigma$ for the difference in sulphur abundance. Therefore, further deep X-ray observations (e.g. with the upcoming XRISM telescope; \cite{Tashiro2020}) are necessary to confirm or deny this hypothetical phenomenon. 

Noteworthily, five background X-ray sources (Table~\ref{XrayBkgPt}) are also identified inside/around the angular dimension of SNR G298.6$-$0.0. \PKHY{All of them are well separated from the centroids of our identified $\gamma$-ray sources ($>2\sigma$), making them unlikely to be major contributors to the detected GeV $\gamma$-rays.} Only two of them (Pt2 and Pt4) have  sufficiently high photon statistics for spectral analyses. Our spectral fittings for Pt2 and Pt4 yield $N_\mathrm{H}$ values that are less than half of the $N_\mathrm{H}$ we obtained for the SNR, implying that Pt2 and Pt4 are  \PKHY{closer to us than  the SNR by a few kpc}. It is, therefore, unreasonable to associate these background X-ray sources with a putative pulsar originated from the same supernova explosion as G298.6$-$0.0. 

Furthermore, both the spectra of Src-NE and Src-NW (Figure~\ref{SEDplot}) follow the simple PL model from 2~GeV to $\gtrsim$30~GeV, in contrast with typical spectra of pulsars having an exponential cutoff at about 1--4~GeV \PKHY{(at most 10~GeV for some extreme cases; \cite{Abdo2013})}. It is, therefore, safe to exclude the possibility that the $\gamma$-ray emission of Src-NE or Src-NW is originated from a putative pulsar. 

\subsection{Interactions of SNR G298.6$-$0.0 with MCs}\label{hadronic_SNR--MC}

We argue that the cavity-like structure in the $p-v$ diagram of CO represents an expanding gas motion likely caused by the supernova shocks and/or strong stellar winds from the progenitor of the SNR (e.g., \cite{Koo1990, Koo1991, Hachisu1996, Hachisu1999a, Hachisu1999b}). It is noteworthy that the apparent size of such a wind-cavity is expected to be the same as that of the SNR because the free expansion phase inside the wind-cavity is very short owing to a much lower density (e.g., \cite{Weaver1977}). Moreover, the expansion velocity of the wind-cavity in SNR G298.6$-$0.0 is $\sim$8~km~s$^{-1}$, which is consistent with those in Galactic/Magellanic SNRs ($\sim$3--13~km~s$^{-1}$; e.g., \cite{Landecker1989, Fukui2012, Kuriki2018, Sano2019, Sano2021a, Sano2021b}). 

The physical association between the CO cavity and the SNR is further supported by the HI absorption. The absorption feature is significant only in the $V_\mathrm{LSR}$ range of $-$46--+32~km~s$^{-1}$, whose positive-$V_\mathrm{LSR}$ end is roughly equal to the maximum $V_\mathrm{LSR}$ of the CO cavity. We, therefore, propose  that the wind-cavity at $V_\mathrm{LSR}$ = 19--34~km~s$^{-1}$ was formed by the strong stellar winds from the high-mass progenitor and is now physically interacting with the supernova shocks.

According to the Galactic rotation curve model \citep{Brand1993} with the IAU-recommended values of $R_0$=8.5~kpc and $\Theta_0$=220~km~s$^{-1}$ \citep{Kerr1986}, the $V_\mathrm{LSR}$ of 27$\pm$8~km~s$^{-1}$ harbouring the cavity of the CO expanding shell corresponds to a kinematic distance of 10.1$\pm$0.5~kpc between us and the SNR. This value of kinematic distance is in good agreements with the maximum distance of us to most interstellar hydrogen in this direction (10~kpc; see the face-on maps of \cite{Nakanishi2006}), as well as a measurement based on the ``radio surface brightness vs. diameter" relation (9.5~kpc; \cite{Case1998}).  

At $V_\mathrm{LSR}\sim19$~km~s$^{-1}$ along our line of sight to the SNR  (i.e. slightly in front of the CO cavity), there is an MC clump traced by the CO $p-v$ diagram. This clump is slightly more concentrated in the northern half of the SNR's angular dimension, as shown by the integrated-intensity map.  Right next to the eastern edge of the SNR, the CO emissions reveal a larger and denser MC clump centered at $V_\mathrm{LSR}\sim22$~km~s$^{-1}$. In a scenario of SNR--MC interaction, \PKHY{the combined contribution of these two MC clumps could possibly account for the $\gamma$-ray emission of Src-NE, whose 95\% error circle of the centroid significantly overlaps with both of them}. 

Along our line of sight to the western vicinity of the SNR, the CO emissions also reveal another dense MC clump, which is centered at $V_\mathrm{LSR}\sim29$~km~s$^{-1}$ and is spatially coincident with Src-NW. This clump is also potentially interacting with the SNR, and hence could possibly explain the $\gamma$-rays of Src-NW \PKHY{whose 95\% error circle of the centroid encloses most of its CO emission}. 

$\gamma$-ray sources of SNR--MC hadronic interaction detected by \emph{Fermi}-LAT commonly have spectral breaks in the GeV band \citep{Acero2016}. Src-NE has a spectral break at $E_{br}=1.83\pm0.44_{stat}{^{+0.20}_{-0.09}}_{sys}$~GeV, further supporting an emission mechanism where  MCs are impacted by  SNR-accelerated hadronic cosmic-rays. The break energy of the Src-NW spectrum is constrained to be $\lesssim$1.2~GeV \PKHY{according to the flux bins. Despite an undetermined lower limit for the break energy,  a hadronic scenario of an SNR--MC system is a viable option for interpreting the Src-NW emission as well. }

Our results of $\gamma$-ray, X-ray and radio data analyses unanimously point to the same implication that the interaction of G298.6$-$0.0 with MCs occurs mainly in the northeast direction. Our observational evidences for this argument are as follows:  (i) The $>$2~GeV integrated flux of Src-NE is much greater than  Src-NW's (Table~\ref{2GeVSpecPara});  (ii) The extended keV emission slightly slants to the northeast part of the SNR, contrasting harshly with the radio-continuum morphology \citep{Whiteoak1996} which has the brightest shell at the southwest edge;  (iii) In the vicinity of the SNR, the MCs (traced by CO) on the north, northeast and east sides are larger, denser and closer to the SNR. 

We note that the neutral iron line at $\sim$6.4~keV can be emitted from SNR regions, due to the SNR-accelerated MeV cosmic-ray protons ionising the iron in MCs \citep{Nobukawa2018, Nobukawa2019}. Hence, this line emission can be treated as another tracer of SNR--MC interaction. Unfortunately, Morikawa et al. (in prep.) reports a non-detection of this emission line from our targeted SNR G298.6$-$0.0. Deeper observations in the future are needed to unveil the properties of the neutral iron line in G298.6$-$0.0.

\subsection{Estimated old age of SNR G298.6$-$0.0}\label{oldage}

\PKHY{For Src-NE and Src-NW, the photon indices above the break energies are rather steep (about 2.5--2.8), suggesting that these spectral breaks are associated with the cosmic-ray escape from SNR G298.6$-$0.0 and hence could hint at the SNR age. }

\citet{Suzuki2022} explored the relation between the break energy of an SNR-associated $\gamma$-ray source and the SNR age (see Figure~2(b) of that paper). Based on a sample of $\sim$20 SNRs with ages of $\sim$(0.34--69)~kyr \PKHY{and break energies of $\sim$(3~GeV--1~TeV)}, they confirm a trend that the break energy decreases with the age. This observed trend agrees with the theory predicting a gradual decline of the escape energy of SNR-accelerated cosmic-rays with time \citep{Ptuskin2003, Ptuskin2005}. \citet{Suzuki2022} approximate the $\gamma$-ray break energy as a power-law function of the SNR age. If we substitute $E_{br}\sim1.8$~GeV of Src-NE or $E_{br}\lesssim1.2$~GeV of Src-NW into \PKHY{the power law with the best-fit parameters}, then the age of SNR G298.6$-$0.0 would be computed to be \PKHY{an unrealistically large value for a GeV-bright SNR. Taking into account the large uncertainties of the power-law parameters, an SNR with a given break energy $E_{br}$ should be at least older than $(\frac{E_{br}}{140~\mathrm{GeV}})^{-0.68}$~kyr. Considering the statistical and systematic uncertainties of $E_{br}$ of Src-NE, its upper limit is, at the very most, 3.3~GeV.} We hereby place a conservative and reliable constraint of $>$10~kyr on the age of G298.6$-$0.0.

It is noteworthy that an old ($\gtrsim$10~kyr) SNR may have a broken shell formed from the shock-cloud interaction. An additional component of lower-energy cosmic-rays may be leaked from this broken shell, neglecting the escape energy required for runaway cosmic-rays. This additional component could lead to a more dramatic drop in $E_{br}$ of the $\gamma$-ray spectrum. For instance, this scenario has been applied to interpret the $E_{br}\sim1.0$~GeV \citep{Cui2018} associated with $>$30~kyr SNR W28 \citep{Velazquez2002}, and the $E_{br}<2$~GeV \citep{He2022} associated with 8.3$\pm$0.5~kyr SNR Kes 79 \citep{Kuriki2018}. Besides, simulations (e.g., \cite{Lee2015, Yasuda2019, Kobashi2022}) revealed a wide diversity of evolutionary routes of the $\gamma$-ray emission strongly correlated with the environmental characteristics of an SNR, such as the density distribution of ambient gas and the profile of magnetic field. This diversity could naturally entail a considerable range of possible ages for a given $E_{br}$ in $\gamma$-ray.

\PKHY{The environmental diversity and/or the broken-shell-leaked cosmic-ray  could explain why the observed $E_{br}$ values of the outlying  SNRs in the sample of \citet{Suzuki2022} appear to be lower than their power-law predictions. Among these outlying SNRs, the two with the lowest break energies ($\sim$3~GeV) are about 11--13~kyr old. Thus, our lower limit of $>$10~kyr on the age of G298.6$-$0.0 would remain valid even if G298.6$-$0.0 is an outlying SNR. }

The physical size of SNR G298.6$-$0.0 can provide an independent constraint on its age. In \S\ref{hadronic_SNR--MC}, we derive a kinematic distance of 10.1$\pm$0.5~kpc between us and G298.6$-$0.0. As observed by MOST at the radio frequency of 843~MHz \citep{Whiteoak1996}, this SNR has an angular radius of $\sim5.3^{'}$, which corresponds to a physical radius of $\sim$15.5~pc at 10.1~kpc from us. Such a large physical size also suggests an old SNR age of $>$10~kyr (see Figure 7 of \cite{Cui2018}).

Moreover, our Chandra-ACIS X-ray spectra reconstructed for the inner and whole regions of the diffuse emission, as well as the Suzaku X-ray spectrum of \citet{Bamba2016}, can both be fit with the $\mathrm{VAPEC{\times}PHABS}$ model, which assumes a collisional ionisation equilibrium of the thermal plasma. Such good fits imply that G298.6$-$0.0 had already passed the long-lasting ionization-dominant stage (duration $>$ 10~kyr; e.g., \cite{Masai1984, Smith2010}), further supporting the old age estimated from the $\gamma$-ray spectra and the radio SNR size.

\subsection{Caveats: Systematic effects of background modelings on \emph{Fermi}-LAT spectral analyses} \label{caveat}

In the $\gamma$-ray spectral energy distribution of Src-NE (Figure~\ref{SEDplot}(a)), the partial discrepancy between the BKPL model and the binned spectrum deserves our attention. Below the spectral break of $E_{br}\sim1.8$~GeV, the flux measurements for those individual bins are collectively lower than the predictions by BKPL, and the binned spectrum has an even harder photon index than the best-fit $\Gamma_1$ of BKPL. In  the Galactic diffuse model (gll\_iem\_v07.fits), we notice that the spatial distribution of the modelled diffuse $\gamma$-rays at/around Src-NE is very similar to the MC distribution traced by our CO map (Figure~\ref{G298_CO_pv}(a)). In other words, the Galactic diffuse component significantly overlaps with the PSF of Src-NE. We speculate that the discrepancy between the model line and individual bins at lower energies is owing to the inaccurate energy-dependence of gll\_iem\_v07.fits for the Src-NE region. 

On the other hand, it is comforting to note two things: (i) the apparent difference in lower-energy photon index between the line and bins is consistent with our quantified systematic uncertainty of $\Gamma_1$ of BKPL (Table~\ref{FermiSpecPara}); \PKHY{(ii) the flux difference between the model line and each bin below the break energy is within the quadratic combination of statistical and systematic uncertainties}. We also stress that our detection of the spectral break is robust against the combination of statistical and systematic uncertainties.

The CO MC clumps at/around another $\gamma$-ray source Src-NW appear to be omitted from the model of Galactic diffuse $\gamma$-rays. With regards to this issue, we recommend that the origins of Src-NW should be interpreted as the Galactic cosmic-ray sea in addition to the cosmic rays accelerated by SNR G298.6$-$0.0.

In a more detailed modeling approach, it is crucial to distinguish the physical meanings derived from different $\Gamma_1$ values of the Src-NE spectrum. \PKHY{To be specific, the relatively soft index $\Gamma_1\sim1.9$ yielded by the BKPL fit would imply there is another break at a lower energy which is the signature of pion-decays, while the harder $\Gamma_1$ demonstrated by the binned spectrum would suggest a sharp (i.e. quasi-monoenergetic) proton spectrum.} It is also crucial to constrain the relative contributions of SNR G298.6$-$0.0 and the Galactic cosmic-ray sea to the observed Src-NW emission.

\section*{Acknowledgments}

The NANTEN project is based on a mutual agreement between Nagoya University and the Carnegie Institution of Washington (CIW). We greatly appreciate the hospitality of all the staff members of the Las Campanas Observatory of CIW. We are thankful to many Japanese public donors and companies who contributed to the realization of the project. The Australia Telescope Compact Array (ATCA) and the Parkes radio telescope are parts of the Australia Telescope National Facility which is funded by the Australian Government for operation as a National Facility managed by CSIRO. 
PKHY thanks the Japan Society for the Promotion of Science (JSPS) fellowship (id. PE21024). This work is also supported in part by Grants-in-Aid for Scientific Research from the Japanese Ministry of Education, Culture, Sports, Science and Technology (MEXT) of Japan, No. 19K03908 (AB) and 21H01136 (HS). We thank S.-H. Lee of Kyoto University, T. Mizuno of Hiroshima University,  Kumiko K. Nobukawa of Kindai University and H. Suzuki of Konan University for useful discussion. We also thank the anonymous referee for the fruitful comments.

%%%%%\bibliography{Zotero}
%%%%%\bibliographystyle{apj}

\end{document}